\documentclass[english,aps,groupedaddress,superscriptaddress,nofootinbib]{revtex4}
\usepackage{amsfonts}
\usepackage{amssymb}
\usepackage{amsmath}
\usepackage{color}
\usepackage{dcolumn}
\usepackage{bm}
\usepackage{babel}

\usepackage{amsmath}
\usepackage{graphicx}
\usepackage{amsfonts}
\usepackage{amssymb}
\usepackage{hyperref}
\usepackage{color}
\usepackage{mathrsfs}
\usepackage{isomath}
\usepackage{amsmath}
\usepackage{amsthm}
\usepackage{epstopdf}
\usepackage{txfonts}
\usepackage{float}
\usepackage{dsfont}

\newcommand{\ket}[1]{|#1\rangle}

\newcommand{\Tr}{\mbox{Tr}}

\def\be{\begin{equation}}
\def\ee{\end{equation}}
\def\ba{\begin{array}}
\def\ea{\end{array}}
\def\Tr{\mathrm{Tr}}

\newtheorem{thm}{Theorem}

\newtheorem{cor}[thm]{Corollary}

\begin{document}

\title{$\mbox{Quasi-Fine-Grained Uncertainty Relations}$}

\author{Yunlong~Xiao}
\thanks{Y. Xiao and Y. Xiang contributed equally.}
\address{Department of Mathematics and Statistics, University of Calgary, Calgary, Alberta T2N 1N4, Canada}
\address{Institute for Quantum Science and Technology, University of Calgary, Calgary, Alberta, T2N 1N4, Canada}
\author{Yu~Xiang}
\thanks{Y. Xiao and Y. Xiang contributed equally.}
\affiliation{State Key Laboratory of Mesoscopic Physics, School of Physics, Nano-optoelectronics Frontier Center of the Ministry of Education $\&$ Collaborative Innovation Center of Quantum Matter, Peking University, Beijing 100871, China}
\affiliation{Beijing Academy of Quantum Information Sciences, Beijing 100193, China}
\affiliation{Collaborative Innovation Center of Extreme Optics, Shanxi University, Taiyuan, Shanxi 030006, China}
\author{Qiongyi~He}
\email{qiongyihe@pku.edu.cn}
\affiliation{State Key Laboratory of Mesoscopic Physics, School of Physics, Nano-optoelectronics Frontier Center of the Ministry of Education $\&$ Collaborative Innovation Center of Quantum Matter, Peking University, Beijing 100871, China}
\affiliation{Beijing Academy of Quantum Information Sciences, Beijing 100193, China}
\affiliation{Collaborative Innovation Center of Extreme Optics, Shanxi University, Taiyuan, Shanxi 030006, China}
\author{Barry C. Sanders}
\email{sandersb@ucalgary.ca}
\address{Institute for Quantum Science and Technology, University of Calgary, Calgary, Alberta, T2N 1N4, Canada}

\begin{abstract}
{\bf Keywords:} Uncertainty Principle, Quantum Memory, Nonlocality, Einstein-Podolsky-Rosen steering
\end{abstract}

\maketitle

\section*{Abstract}
Nonlocality, which is the key feature of quantum theory, has been linked with the uncertainty principle by fine-grained uncertainty relations, by considering combinations of outcomes for different measurements. However, this approach assumes that information about the system to be fine-grained is local, and does not present an explicitly computable bound. Here, we generalize above approach to general quasi-fine-grained uncertainty relations (QFGURs) which applies in the presence of quantum memory and provides conspicuously computable bounds to quantitatively link the uncertainty to entanglement and Einstein-Podolsky-Rosen (EPR) steering, respectively. Moreover, our QFGURs provide a framework to unify three important forms of uncertainty relations, i.e., universal uncertainty relations, uncertainty principle in the presence of quantum memory, and fine-grained uncertainty relation. This result gives a direct significance to the uncertainty principle, and allows us to determine whether a quantum measurement exhibits typical quantum correlations, meanwhile, it reveals a fundamental connection between basic elements of quantum theory, specifically, uncertainty measures, combined outcomes for different measurements, quantum memory, entanglement and EPR steering.

\section{Introduction} 
The uncertainty principle, articulated in 1927 by Heisenberg~\cite{Heisenberg1927}, plays a crucial role in highlighting the non-classical nature of quantum probabilities. It states that the outcomes of two incompatible measurements cannot be predicted simultaneously with certainty. The formal inequality based on the standard deviations of position and momentum was derived by Kennard~\cite{Kennard1927} and Weyl~\cite{Weyl1927}, and generalized by Robertson~\cite{Robertson1929} and Schr\"{o}dinger~\cite{Schrodinger1930} for general observables. Even though variance-based uncertainty relations play an important role in quantum theory \cite{Huang2012, Maccone2014, Xiao2016W, Xiao2017I, Maccone2018}, later information-theoretic entropy was introduced as a natural way to quantify uncertainty, and the entropic formulation of uncertainty relations for quantum measurements was widely studied~~\cite{Bialynicki1975, Deutsch1983, Partovi1983,Kraus1987,Maassen1988,Ivanovic1992,Sanchez1993,Ballester2007,Wu2009,Partovi2011,Huang2011,Tomamichel2011,Coles2012,Coles2014, Xiao2016S, Xiao2016QM, Xiao2016U,XiaoPhD,Xiao2019CIP}. However, to fully capture the essence of uncertainty, uncertainty measures in the strictest sense must be monotonically nondecreasing under two classes: randomly chosen symmetry transformations ($\mathcal{D}^{\text{sym}}$) and classical processing via channels followed by recovery ($\mathcal{D}^{\text{rec}}$)~\cite{Narasimhachar2016}. In these cases, nonnegative Schur-concave functions~\cite{Schur} are qualified candidates for uncertainty measures, and one can build various uncertainty relations based on majorization relations and nonnegative Schur-concave functions. Moreover, based on the form of joint uncertainty \cite{Yuan2019,Yuan2019S}, majorization uncertainty relations can be divided into two major categories: direct-product majorization uncertainty relations, i.e. universal uncertainty relations (UURs)~\cite{Friedland2013, Puchala2013}, and direct-sum majorization uncertainty relations~\cite{Rudnicki2014}. 

A significant application of uncertainty relations is to determine the degree of nonlocality \cite{Pramanik2013, Jia2017, Maccone2018M,Wang2018}, which gives the link to security for quantum cryptography~\cite{Coles2017E}. For instance, Berta et al.~\cite{Berta2010} derived the uncertainty principle in the presence of quantum memory (UPQM), and provided a lower bound on the uncertainty denoted by conditional von Neumann entropy corresponding to the measurements on the system $A$ given information stored in the system $B$ (i.e., quantum memory) \cite{Brennen2015}. This bound depends on the degree of entanglement between $A$ and $B$. Oppenheim and Wehner~\cite{Oppenheim2010} demonstrated the quantitative connection between uncertainty and nonlocality of quantum games by applying a fine-grained uncertainty relation (FGUR), showing that the amount of nonlocality can determine the strength of uncertainty in measurements. 

Each of UURs, UPQM, and FGUR captures different features of uncertainty. Specifically, UURs contain the diversity of uncertainty measures, the UPQM links the uncertainty to the amount of quantum entanglement between subsystems by measuring entropic functions, while FGUR consists of a series of inequalities to include all possible combinations of outcomes for different measurements. The question, thus, naturally arises: can all these uncertainty relations be unified into a general form? On the other hand, even though FGUR nicely presents a close connection between a quantum game in terms of Bell inequalities and uncertainty relation based on the winning condition (a particular choice of measurement outcomes), one would be curious can different aspects of quantum correlations, such as Einstein-Podolsky-Rosen (EPR) steering~\cite{Einstein1935,Howard07PRL} and entanglement be also linked with uncertainty from a single framework with clearly computable bounds of FGURs. We answer these questions in the affirmative by revisiting Schr\"odinger's concept of probability relations between separated systems \cite{Schrodinger1935, Schrodinger1936} from a quantum information perspective. 

\begin{figure*}[t]
\begin{center}
\includegraphics[width=180mm]{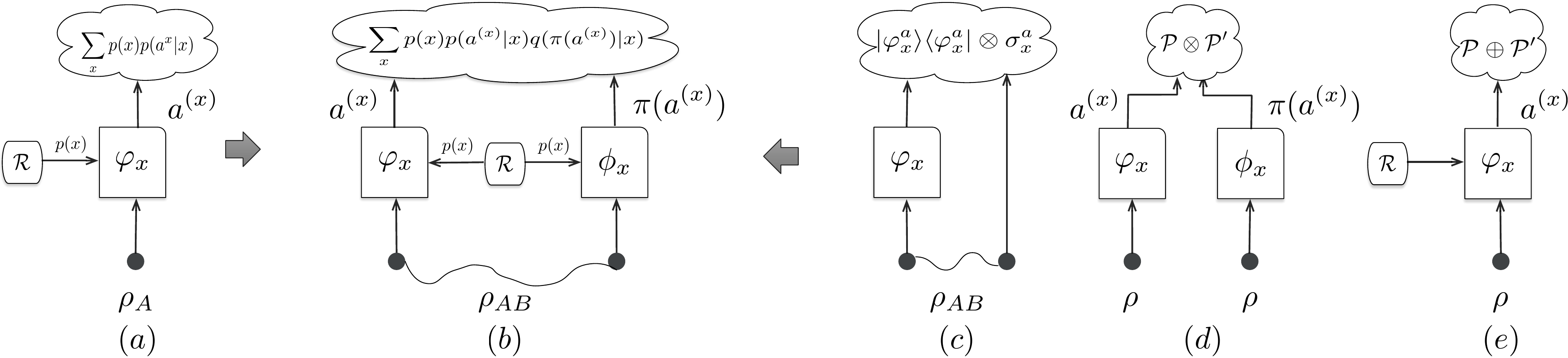}
\end{center}
\caption{Process diagrams for uncertainty relations. $R$ denotes random number generator. (a) Fine-grained uncertainty relation (FGUR)~\cite{Oppenheim2010}; (b) Our Quasi-Fine-Grained Uncertainty Relations (QFGURs); (c) Uncertainty principle in the presence of quantum memory (UPQM)~\cite{Berta2010}; (d) Universal uncertainty relations (UURs)~\cite{Friedland2013, Puchala2013}; (e) Direct-sum majorization uncertainty relations~\cite{Rudnicki2014}.}
\label{ur}
\end{figure*}
 
Here we introduce a simple but universally applicable theory of quantum probability relations (QPRs), namely, quasi-fine-grained uncertainty relations (QFGURs), which unifies UURs, UPQM, and FGUR. In addition, the framework of QFGURs allows us to formulate bounds for both EPR steering and entanglement, which quantitatively connects uncertainty principle to different aspects of nonlocality. EPR steering is defined in terms of violations of a local hidden state (LHS) model and describes the ability that local measurements on one subsystem can remotely pilot (steer) the state of another subsystem~\cite{Einstein1935,Howard07PRL}.
Our theory is based on the notion of local probability relations from a measured system $A$ and a quantum memory $B$, which obey QPRs. Summing overall outcomes for each measurement in QPRs helps to study the unbounded violation of both quantum steering and entanglement inequalities systematically and efficiently. Moreover, our methods clarify that the LHS model itself can be formulated in terms of incompatibility of the available local observables, which sheds light on understanding the intrinsic asymmetry of EPR steering. For illustrative purposes of the general framework, we provide a numerical example and show that our approach actually gives a lower bound to test steering and entanglement than previous coarse-graining entropic functions.

\section{Preliminaries}
First, we generalize Schr\"odinger's discussion of probability relations~\cite{Schrodinger1935, Schrodinger1936} from a quantum information perspective for a task: start with our two protagonists, Alice and Bob. Alice prepares a bipartite quantum state $\rho_{AB}$, holds subsystem $A$ and transmits subsystem $B$ to Bob. This process can be repeated as many times as required. In each round, they measure their system and communicate classically. Alice chooses one of her measurement settings $x\in \mathbb{N}_{N}$, where $\mathbb{N}_{N}:=\left\{1, \ldots,  N\right\}$. She then measures a nondegenerate observable $A_{x}$ with eigenvectors $\left\{\varphi_{x}^{a}\right\}$, and receives an outcome $a^{(x)}$ with probability $p\left(a^{(x)}|x\right) := \Tr\left\{(|\varphi_{x}^{a}\rangle\langle\varphi_{x}^{a}|\otimes \mathds{1})\rho_{AB}\right\}$, where $\mathds{1}$ denotes the identity matrix. The corresponding notations for Bob are $B_{y}$, $\{\phi_{y}^{b}\}$, $b^{(y)}$, and $q\left(b^{(y)}|y\right)$.

We are interested in a general case, where Alice and Bob choose measurements $A_{x}$, $B_{y}$ according to some joint distribution $p(x, y)$. Then, for each combination of possible outcomes $\bm{a}=\left(a^{(1)},a^{(2)},\ldots,a^{(N)}\right)$ and $\bm{b}=\left(b^{(1)},b^{(2)},\ldots,b^{(N)}\right)$ for a fixed set of measurements $x$ and $y$, we define the following QPRs comprising a series of inequalities  
\begin{equation}\label{QPR}
U_{\text{QPR}}:=\left\{\sum\limits_{x, y=1}^{N}p\left(x, y\right) p\left(a^{(x)}|x\right) q\left(b^{(y)}|y\right)\leqslant \zeta^{\text{QPR}}_{\left(a, b\right)} | \forall \bm{a}, \bm{b} \right\}.
\end{equation}
Here, the upper bound $\zeta^{\text{QPR}}_{\left(a, b\right)}$ is the maximization taken overall $\bm{a}$, $\bm{b}$ and states $\rho_{AB}$, which restricts the set of allowed probability distributions. To clearly link QPRs with quantum correlations and uncertainty relations, we introduce a special formalization of QPRs, namely, QFGURs. In the following we show that we can unify UURs, UPQM, and FGUR through QFGURs. 

In the FGUR approach~\cite{Oppenheim2010}, Alice has access to an unknown quantum state $\rho_{A}$ and performs measurement $A_x$ with probability $p\left(x\right)$ which is decided by a random number generator $R$, as indicated in Fig.~\ref{ur} (a). In this setting, the probability of obtaining outcome $a^{(x)}$ under measurement $x$ is determined by $p\left(a^{(x)}|x\right)=\Tr\left(|\varphi_{x}^{a}\rangle\langle\varphi_{x}^{a}|\rho_{A}\right)$. In order to gain more insight into the joint uncertainty through a fine-grained way and how the combinations of different measurement outcomes affect the total uncertainties, Oppenheim and Wehner~\cite{Oppenheim2010} considered the following inequalities
\begin{equation}\label{FGUR}
U_{\text{FGUR}}:=\left\{\sum\limits_{x=1}^{N}p\left(x\right) p\left(a^{(x)} | x\right)\leqslant \zeta^{\text{FGUR}}_{a} \, | \, \forall \bm{a}  \right\},
\end{equation}
which only focuses on Alice's measured system, whereas Eq. (\ref{QPR}) considers the probabilities stemming from both Alice and Bob. Here, the value of $\zeta^{\text{FGUR}}_{a}$ is the maximization taken overall quantum state $\rho_A$. To confirm that Eq. (\ref{FGUR}) forms a joint uncertainty, we only need to check whether $\zeta^{\text{FGUR}}_{a}<1$ is satisfied, as it prescribes that one cannot obtain deterministic forecast whenever $\sum_{x}p\left(x\right) p\left(a^{(x)} | x\right)<1$.

\section{Quasi-Fine-Grained Uncertainty Relations (QFGURs)}

Instead of performing measurements on a single system, now we consider local measurements
$\left\{\varphi_{x}\right\}$ and $\left\{\phi_{x}\right\}$, as delineated in Fig.~\ref{ur} (b), on a bipartite quantum system $\rho_{AB}$ that has two subsystems of the same dimensionality. In this case, the measure of uncertainty of outcomes for the measured system is conditioned on the information in quantum memory stored in the other system~\cite{Berta2010}. When Alice obtains a result $a^{(x)}$ with probability $p\left(a^{(x)}|x\right)$, the conditional quantum memory $\sigma_{x}^{a}=\Tr_{A}\left\{(|\varphi_{x}^{a}\rangle\langle\varphi_{x}^{a}|\otimes \mathds{1})\rho_{AB}\right\}/\Tr\left\{(|\varphi_{x}^{a}\rangle\langle\varphi_{x}^{a}|\otimes \mathds{1})\rho_{AB}\right\}$, renormalized to have unit trace, is created some distance away at Bob's location. Bob's resulting probability to obtain a result $\pi(a^{(x)})$ for his observable $B_{x}$ with eigenvectors $\left\{\phi_{x}^{\pi(a)}\right\}$ is thus quantified by $q\left(\pi(a^{(x)})|x\right)=\Tr\left(|\phi_{x}^{\pi(a)}\rangle\langle\phi_{x}^{\pi(a)}|\sigma_{x}^{a}\right)$,  which depends on the choice of measurement $x$, the corresponding outcome $a^{(x)}$ from subsystem $A$ and the permutation $\pi$ performed on subsystem $B$.

As discussed previously, the first step in the fine-grained approach is to identify a simple sorting method which is realized by a random number generator $R$ between Alice and Bob. Here we chose to take measurement with index $x$, i.e. $\left\{\varphi_{x}\right\}$ on subsystem $A$ and then $\left\{\phi_{x}\right\}$ on subsystem $B$, according to the probability $p(x)$ which is setted by $R$ as shown in Fig.~\ref{ur}(b). From this figure, we can also identify the features of above discussion as the following set of inequalities
\begin{equation}\label{QUR}
U_{\text{QFGUR}}:=\left\{\sum\limits_{x=1}^{N}p\left(x\right) p\left(a^{(x)}| x\right) q\left(\pi\left(a^{(x)}\right) | x\right)\leqslant \zeta^{\text{QFGUR}}_{a}|\forall \textbf{a}, \pi \right\},
\end{equation}
which forms QFGURs. Here, each measurement $x$ can result in one of $d$ possible outcomes, i.e. $a^{(x)}\in \mathbb{N}_{d}$, and $\pi \in \mathfrak{S}_{d}$ can be any permutation of the outcomes with symmetric group $\mathfrak{S}_{d}$. Notably, the upper bound $\zeta^{\text{QFGUR}}_{a}$ is a function of the combination of possible outcomes $\bm{a}$, which allows us to derive the bound by maximizing overall quantum states $\rho_{AB}$ and permutations $\pi$. Named ``Quasi'' means the distribution relaxes the Kolmogorov's axioms of probability theory, i.e., $\sum^{d}_{a^{(x)}=1} \bar{p}\left(a^{(x)}| x\right)<1$ in general, by setting $\bar{p}(a^{(x)} | x):=p\left(a^{(x)}| x\right) q\left(\pi\left(a^{(x)}\right) | x)\right)$ (Actually, the raise of Kolmogorov's axiomatic probability theory provided a solution to the sixth Hilbert problem~\cite{Probability}).

\subsection{The Unification of Various Forms of Uncertainty Relations}

Consequently, UPQM in terms of entropic functions can be obtained directly from QFGUR. To do so, we remove the random number generator $R$ and do nothing on subsystem $B$. We can now express the result state as $|\varphi_{x}^{a}\rangle\langle\varphi_{x}^{a}|\otimes\sigma_{x}^{a}$, as shown in Fig.~\ref{ur}(c), which leads to UPQM by applying Shannon entropy. In arriving at FGURs, we simply neglect subsystem $B$.

Our discussion so far focus on developing the general formalism of QFGUR by assuming local operations on subsystems and shared randomness between them, which is related to UURs. Of course, one can pick $\rho_{AB}=\rho\otimes\rho$ and remove the shared randomness $R$ between them to formulate the joint uncertainty via direct-product, i.e. UURs. As portrayed in Fig.~\ref{ur}(d), by measuring identically prepared quantum state $\rho$ with respect to bases $\left\{ |\varphi_{x}^{a}\rangle\langle\varphi_{x}^{a}| \right\}$ and $\left\{ |\phi_{x}^{a}\rangle\langle\phi_{x}^{a}| \right\}$, their probabilities can be collected as $\bm{\mathcal{P}}$ and $\bm{\mathcal{P'}}$, respectively. Whenever $\bm{\mathcal{P}} \otimes \bm{\mathcal{P'}} \nprec (1, 0, \ldots, 0)$ with $\prec$ standing for majorization~\cite{Schur}, we cannot predict the measurement outcomes with certainty. As a bonus, the direct-sum majorization uncertainty relations can also be obtained from QFGURs by ignoring one subsystem and then performing measurements with uniform probability distribution according to the random number generator $R$, as shown in Fig.~\ref{ur}(e). All rigorous mathematical proofs are given in the appendices.

\subsection{Witness for Entanglement and EPR steering}   

For simplicity, we now characterize the amount of uncertainty in a physical system while taking a particular permutation $\pi=(1) \in \mathfrak{S}_{d}$. We are interested in the values of the upper bound $\zeta^{\text{QFGUR}}_{a}\left(\mathcal{M}\right)=\max_{\rho_{AB\in \mathcal{M}}}\left\{\sum_{x}p(x) p(a^{(x)} | x) q(a^{(x)} | x)\right\}$, where the maximization is taken overall states within a specific type of quantum states $\rho_{AB}\in\mathcal{M}$. Here, $\mathcal{M}$ can be any quantum states ($\mathcal{Q}$), separable states ($\mathcal{S}$), or the bipartite states allowed for LHS model ($\mathcal{E}$), and the hierarchy relations of those states are sketched in $\zeta_{a}\left(\mathcal{S}\right) \leqslant \zeta_{a}\left(\mathcal{E}\right) \leqslant \zeta_{a}\left(\mathcal{Q}\right)$. Furthermore, if a state $\rho_{AB}$ satisfies $\left(\sum_{x}p(x) p(a^{(x)} | x) q(a^{(x)}| x)\right)_{\rho_{AB}}>\zeta_{a}\left(\mathcal{S}\right)$, then $\rho_{AB}$ must be entangled. Hence, from our QFGUR, we construct criteria to test entangled states, steerable states (here we always assume steerable of the type ``$A$ steers $B$''). 

%
%

%
%

Next, we derive a general form of inequalities for entanglement and steering, and provide some numerical examples to show the improvement of their experimentally feasible unbounded violation.

The following few paragraphs consider the quantum EPR steering scenario \cite{Einstein1935, Howard07PRL, shanxiCluster, OneWayNatPhot, OneWayPryde, OneWayGuo, 1sDIQKD_howard, CV-QKDexp, HowardOptica, ANUexp, GiannisQSS, YuQSS}. We use the same notations as appeared in our uncertainty scenario for introducing $U_{\text{QFGUR}}$, where $A$ denotes as a measured system, and regarding Bob's system as quantum memory. When Alice obtains result $a$ for measurement $x$, the conditional quantum memory at Bob's place becomes $\sigma_{x}^{a}$. The information encoded in conditional quantum memory can be quantified by the quantum functional~\cite{Rutkowski2017}, $S_{\mathcal{Q}} :=\sum_{x=1}^{N} \sum_{a=1}^{d} \Tr\left(|\phi_{x}^{a}\rangle\langle\phi_{x}^{a}|\sigma_{x}^{a}\right)$, where the maximal value of $S_{\mathcal{Q}}$ equals $N$ when the bipartite quantum state $\rho_{AB}$ is maximally entangled and measurements are ideal in mutually unbiased bases.

Within the LHS model, Alice performs a measurement $x$ with an untrusted device, and announces the outcome $a$ with probability $p_{\lambda} \left(a | x\right)$ involving the local hidden variable $\lambda$. $\lambda$ represents the possible values a shared classical variable $\Lambda$, also named shared randomness, distributed with the density function $p\left(\lambda\right)$. Now Bob's unnormalized conditional quantum memory is given by $\tau_{x}^{a} = \sum_{\lambda \in \Lambda} p\left(\lambda\right) p_{\lambda} \left(a | x\right) \sigma_{\lambda}$, where $\sum_{\lambda\in\Lambda} p\left(\lambda\right)=1$ and $\sigma_{\lambda}$ is Bob's local hidden state, renormalized to trace $1$, after Alice's measurement~\cite{Howard07PRL}. Denoting the local response probability function in Bob's subsystem as $q_{\sigma_\lambda} \left(a | x\right)= \Tr \left(| \phi_{x}^{a} \rangle \langle \phi_{x}^{a} | \sigma_{\lambda}\right)$ for the given measurement $\{| \phi_{x}^{a} \rangle \langle \phi_{x}^{a} | \}$, the steering (from $A$ to $B$) functional is written as 
%
\begin{equation}\label{functionalepr}
S_{\mathcal{E}} := \sum_{x=1}^{N} \sum_{a=1}^{d} \Tr\left(| \phi_{x}^{a} \rangle \langle \phi_{x}^{a} | \tau_{x}^{a}\right) = \sum\limits_{\lambda \in \Lambda} \sum\limits_{x=1}^{N} \sum\limits_{a=1}^{d} 
p\left(\lambda\right)p_{\lambda} \left(a | x\right) q_{\sigma_\lambda} \left(a | x\right).
\end{equation}
%

Entanglement or nonseparability is a weaker sort of correlation than steering~\cite{Howard07PRL}. Within the quantum separable model $\rho_{AB} = \sum_{\lambda} p\left(\lambda\right) \rho^{A}_{\lambda} \otimes \rho^{B}_{\lambda}$, where $\rho^{A}_{\lambda}$ and $ \rho^{B}_{\lambda}$ are some quantum states, we can also give the entanglement functional
\begin{equation}\label{functionalent}
S_{\mathcal{S}} = \sum\limits_{\lambda \in \Lambda} \sum\limits_{x=1}^{N} \sum\limits_{a=1}^{d} 
p\left(\lambda\right) p_{\rho^{A}_{\lambda}} \left(a | x\right) q_{\rho^{B}_{\lambda}} \left(a | x\right),
\end{equation}
with $p_{\rho^{A}_{\lambda}} \left(a | x\right) =  \Tr \left(| \varphi_{x}^{a} \rangle \langle \varphi_{x}^{a} | \rho^{A}_{\lambda}\right)$ and $q_{\rho^{B}_{\lambda}} \left(a | x\right) =  \Tr \left(| \phi_{x}^{a} \rangle \langle \phi_{x}^{a} | \rho^{B}_{\lambda}\right)$. Note that the local response function $p_{\rho^{A}_{\lambda}} \left(a | x\right)$  of entanglement functional comes from quantum measurements while for steering functional $p_{\lambda} \left(a | x\right)$ may come from classical measurements. We remark that the maximal values of $S_{\mathcal{E}}$ and $S_{\mathcal{S}}$ rely on the choice of measurements $\left\{ |\varphi_{x}^{a}\rangle\langle\varphi_{x}^{a}| \right\}$ and $\left\{ |\phi_{x}^{a}\rangle\langle\phi_{x}^{a}| \right\}$.

One immediately sees that $S_{\mathcal{E}}$ and $S_{\mathcal{S}}$ are special forms of combination of left hand side of quasi-fine-grained inequalities given in Eq.~(\ref{QUR}). For a given measurement $\{| \phi_{x}^{a} \rangle \langle \phi_{x}^{a} | \}$ on Bob's system, the violation of the following inequalities $S_{\mathcal{Q}}\leq \sup_{\rho_{AB} \in \mathcal{E}} S_{\mathcal{E}}$ and $S_{\mathcal{Q}}\leq \sup_{\rho_{AB} \in\mathcal{S}} S_{\mathcal{S}}$ indicates that the quantum state is steerable and entangled, respectively. The maximal degree of violation of the steering and entanglement inequalities is determined by
\begin{equation}\label{devision}
V_{\mathcal{E}} = \frac{\sup_{\rho_{AB} \in \mathcal{Q}} S_{\mathcal{Q}}}
{\sup_{\rho_{AB} \in \mathcal{E}} S_{\mathcal{E}}}>1,\ \ 
V_{\mathcal{S}} = \frac{\sup_{\rho_{AB} \in \mathcal{Q}} S_{\mathcal{Q}}}{\sup_{\rho_{AB} \in \mathcal{S}} S_{\mathcal{S}}}>1,
\end{equation}
respectively.

We start introducing our main theorems by some notations. Give an arbitrary number of $N$ measurement settings $x$, we denote a set of $d \times (k + 1)$ rectangular matrices $\left(| \varphi^{a_{1}}_{x_{1}} \rangle, \ldots, | \varphi^{a_{k+1}}_{x_{k+1}} \rangle\right)$, where $k=0,\ldots, dN-1$, and define the maximal squares of norms for those matrices as \cite{Rudnicki2014}
\begin{align}\label{definitions}
\mathcal{S}_{k}^{A} & := \max\left\{\sigma_{1}^{2}\left(| \varphi^{a_{1}}_{x_{1}} \rangle, \ldots, | \varphi^{a_{k+1}}_{x_{k+1}} \rangle\right)\right\},\notag\\
\mathcal{S}_{k}^{B} & := \max\left\{\sigma_{1}^{2}\left(| \phi^{a_{1}}_{x_{1}} \rangle, \ldots, | \phi^{a_{k+1}}_{x_{k+1}} \rangle\right)\right\},
\end{align}
where $\sigma_{1} \left(\cdot\right)$ stands for the maximal singular value,  $| \varphi^{a}_{x} \rangle$ ($| \phi^{a}_{x}\rangle$) is the $a$th eigenvectors of measurement $A_{x}$ ($B_{x}$), and the maximum ranges overall possible $\left\{ (a_{1}, x_{1}), (a_{2}, x_{2}) \ldots, (a_{k+1}, x_{k+1}) \right\}$ in the case of $(a_{i}, x_{i}) \neq (a_{j}, x_{j})$ whenever $i \neq j$, here $i, j \in \left\{1, 2, \ldots, k+1\right\}$, $a_{i}\in\left\{1, 2, \ldots, d\right\}$, $x_{i}\in\left\{1, 2, \ldots, N\right\}$. 

Our main theorems are the following.
\begin{thm}\label{thm1}
EPR steering (from $A$ to $B$) functional $S_{\mathcal{E}}$ satisfies
\begin{equation}\label{steerIneq}
\sup\limits_{\rho_{AB} \in \mathcal{E}} S_{\mathcal{E}} \leqslant \mathcal{S}_{N-1}^{B}.
\end{equation}
\end{thm}
\begin{cor}
The maximum violation of the EPR steering functional by quantum states is
\begin{equation}
V_{\mathcal{E}} \geqslant \frac{N}{\mathcal{S}_{N-1}^{B}}.
\end{equation}
\end{cor}
\begin{thm}\label{thm3}
Entanglement functional $S_{\mathcal{S}}$ satisfies
\begin{align}\label{entIneq}
\sup\limits_{\rho_{AB} \in \mathcal{S}} S_{\mathcal{S}} \leqslant  \mathcal{S}^{AB}.
\end{align}
with 
\begin{align}
\mathcal{S}^{AB}:=1 + \left(\mathcal{S}_{1}^{A}-1\right)\left (\mathcal{S}_{1}^{B}-1\right) + \left(\mathcal{S}_{2}^{A} - \mathcal{S}_{1}^{A}\right) \left(\mathcal{S}_{2}^{B} - \mathcal{S}_{1}^{B}\right)
+\cdots+\left(\mathcal{S}_{dN-1}^{A} - \mathcal{S}_{dN-2}^{A}\right) \left(\mathcal{S}_{dN-1}^{B} - \mathcal{S}_{dN-2}^{B}\right).
\end{align}
\end{thm}
\noindent The optimal bound for QFGURs with LHS model is equal to $\sup S_{\mathcal{E}}$ and with separable model is equal to $\sup S_{\mathcal{S}}$, and can be in general hard to compute since they involve optimization problems. Nonetheless, our main theorems provide (weaker) upper-bounds for QFGURs, but which are explicitly computable.
\begin{cor}
The maximum violation of the entanglement functional by quantum states is
\begin{equation}
V_{\mathcal{S}} \geqslant \frac{N}{\mathcal{S}^{AB}}.
\end{equation}
\end{cor}
\noindent
Full proofs of Theorems \ref{thm1} and \ref{thm3} are based on the direct-sum majorization uncertainty relation and detailed in the appendices. Although it is shown that all the discussions in this paper focus on the bipartite system $\rho_{AB}$ with same dimensional subsystems, as shown in the explicit method presented in the appendices, this major restriction can be relaxed.

Recently, the authors of Ref.~\cite{Rutkowski2017} focused on the steering functional $\sup_{\rho_{AB} \in \mathcal{E}} S_{\mathcal{E}}$ and provided an upper bound by means of the Deutsch-Maassen-Uffink's entropic relation~\cite{Maassen1988}
\begin{equation}\label{C2017}
\sup_{\rho_{AB} \in \mathcal{E}} S_{\mathcal{E}} \leqslant1+\sum\limits_{i=1}^{N-1}C_{i},
\end{equation}
where $C_{i}:=\max_{x}C_{x(N+x-i\mod N)}$ and $C_{xy}:=\max_{a, b}|\langle \varphi_{a}^{x}|\phi_{b}^{y}\rangle|$ is the maximal overlap between these observables. Their inequality shows that the unbounded violation of steering inequality depends on the maximal overlap between incompatible measurements. 

Here, our Theorem \ref{thm1} proves that this unbounded violation does not only depend on the maximal overlap, but on all overlaps between incompatible measurements. Moreover, it is also possible to show that their upper bound is improved by applying our QFGURs. We give a simple example to show the improvement, which reveals that our method based on QFGURs is more sufficient and relaxed than the one in Eq.~(\ref{C2017})~\cite{Rutkowski2017}. 

\begin{figure}
\center
\includegraphics[width=0.55\columnwidth]{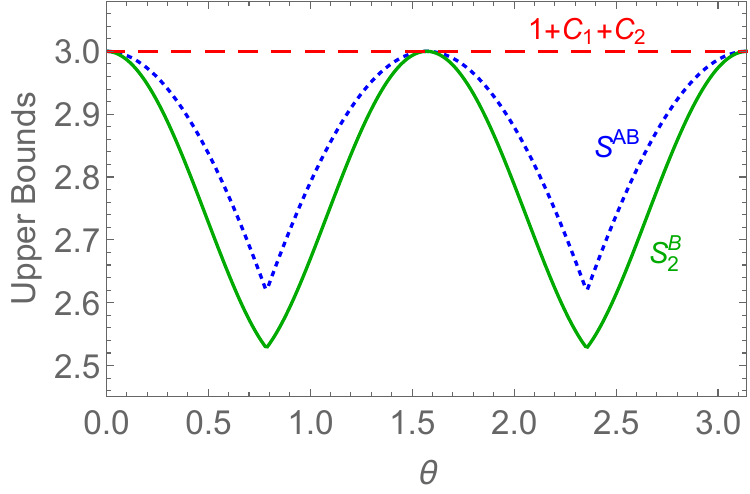}
\caption{For observables considered in Eq.~(\ref{M}), the maximal steering bound $1+C_1+C_2$~\cite{Rutkowski2017}, our steering bound $\mathcal{S}_{2}^{B}$ and entanglement bound $\mathcal{S}^{AB}$.}
\label{3bound}
\end{figure}

Consider the observables $M_{1}, M_{2}$ and $M_{3}$ with the following eigenvectots
\begin{align}\label{M}
M_{1}&=\left\{\begin{pmatrix}1 \\ 0 \\ 0 \end{pmatrix},
\begin{pmatrix}0 \\ 1 \\ 0 \end{pmatrix},
\begin{pmatrix}0 \\ 0 \\ 1 \end{pmatrix}
\right\},~
M_{2}=\left\{\begin{pmatrix}1 \\ 0 \\ 0 \end{pmatrix},
\frac{1}{\sqrt{2}}\begin{pmatrix}0 \\ 1 \\ 1 \end{pmatrix},
\frac{1}{\sqrt{2}}\begin{pmatrix}0 \\ 1 \\ -1 \end{pmatrix}
\right\},~
M_{3}=\left\{\begin{pmatrix}\cos\theta \\ 0 \\ \sin\theta \end{pmatrix},
\begin{pmatrix}0 \\ 1 \\ 0 \end{pmatrix},
\begin{pmatrix}-\sin\theta \\ 0 \\ \cos\theta \end{pmatrix}
\right\}.
\end{align}
Using the uncertainty relations (\ref{steerIneq}), (\ref{entIneq}) and (\ref{C2017}), we obtain three upper bounds $\mathcal{S}_{2}^{B}$, $\mathcal{S}^{AB}$, and $1+\sum^{2}_{i=1}C_i$ to classify steering, entanglement by our method, and steering by the method used in Ref.~\cite{Rutkowski2017}, respectively. As shown in Fig.~\ref{3bound}, it is clear that our method provides bounds tighter than the one constructed in Ref.~\cite{Rutkowski2017} to some extend and any violation of $\mathcal{S}_{2}^{B}$, $\mathcal{S}^{AB}$ implies steerability and entanglement respectively. For Werner states with exactly two and three measurements, our EPR steering criterion (Theorem \ref{thm1}) coincides with the optimal bound $1/\sqrt{2}$ and $1/\sqrt{3}$ Ref.~\cite{Howard07PRL, Cavalcanti2009}, respectively. For the cases with dimensions higher than two, our Theorem \ref{thm1} tends to be no better than other state dependent criteria~\cite{Howard07PRL}, whereas it still performs better than that determined by inequality (\ref{C2017})~\cite{Rutkowski2017}. The examples are detailed in appendices. This is expected since our uncertainty relations is valid for multifarious quantum correlations and is not optimized specifically for EPR steering. In principle, we wish to develop state independent bounds for uncertainty relations which reveal the incompatibility between observables. However, those uncertainties usually do not provide better test for quantum correlations than the state dependent criteria in terms of the eigenvalues of the reduced systems. The challenge is, in order to acquire strong criteria of EPR steering and entanglement which are comparable with other approaches, we need to consider the optimal bound of QFGURs which we would leave for future work.

\section{Conclusions}

We revisit Schr\"odinger's probability relations and sketch process diagrams for the universal uncertainty relations (UURs), uncertainty principle in the presence of quantum memory (UPQM), and fine-grained uncertainty relation (FGUR) in one framework. Based on the notion of local probability relations from a measured system and a quantum memory, we prove a special format, namely a quasi-fine-grained uncertainty relation, which unifies UURs, UPQM, and FGUR. Further, we apply our theory to show that the LHS model itself can be formulated in terms of incompatibility of the available local observables. Thus, we derive the experimentally feasible inequalities to test steering and entanglement, and discuss the unbounded violation, which is not only based on the maximal overlap~\cite{Rutkowski2017} but on all overlaps between incompatible measurements. Here we highlight the role of our framework in tests of entanglement, EPR steering and the uncertainty principle, but it is general and allows us to derive and generalize some results in uncertainty measures. For example, some combinations (like UUR) of our quasi-fine-grained inequalities are monotonic under $\mathcal{D}^{\text{sym}}$ and $\mathcal{D}^{\text{rec}}$~\cite{Narasimhachar2016}, while some are not (like FGUR), their relations were left open and will be addressed elsewhere. Finally, our method paves the way for deeply understanding uncertainty and nonlocality and the fundamental relations between these two striking aspects of quantum mechanics. 

\section*{Acknowledgments}
Discussions with Naihuan Jing and Gilad Gour are gratefully acknowledged. Q.H. thanks the support from the National Natural Science Foundation of China (Grants No. 11622428, No. 11975026, and No. 61675007), Beijing Natural Science Foundation (Grant No. Z190005), and the Key R$\&$D Program of Guangdong Province (Grant No. 2018B030329001). Y.X. acknowledges the National Postdoctoral Program for Innovative Talents (BX20180015), and China Postdoctoral Science Foundation (2019M650291). B.C.S. acknowledges NSERC support.

\appendix*
\setcounter{equation}{0}

\section*{Appendices}
\subsection*{Appendix A: \; QFGURs $\Rightarrow$ UPQM}

To confirm that our QFGURs provide enough information to formulate afresh UPQM, we have investigated
the result state associated with the density matrix $|\varphi_{x}^{a}\rangle\langle\varphi_{x}^{a}|\otimes\sigma_{x}^{a}$ with probability $p\left(a^{(x)}| x\right)$ when the random number generator $R$ is absent while measurement $\varphi_{x}$ is only performed on subsystem $A$. Clearly, the uncertainties of the measured system $A$ and the quantum memory $B$ are determined by
\begin{align}
H(\varphi_{x}, B) := H\left(\sum\limits_{a} p\left(a^{(x)}| x\right) |\varphi_{x}^{a}\rangle\langle\varphi_{x}^{a}|\otimes\sigma_{x}^{a}\right),
\end{align}
where we use $H$ to denote von Neumann entropy. Note that in this appendix, no distinction in notation of von Neumann entropy and Shannon entropy is made, and both would be represented by the same symbol $H$. Whenever the object of function $H$ is a probability distribution, $H$ stands for Shannon entropy; otherwise $H$ represents von Neumann entropy. Thus, the uncertainties conditioned on quantum memory $B$ are
\begin{align}
H(\varphi_{x}|B) := H(\varphi_{x}, B) - H(B),
\end{align}
where $H(B) := H(\rho_{B})$ is the information gained from measuring the quantum memory $B$. Under these conditions, the joint uncertainties between measurements $\varphi_{x}$ and $\phi_{x}$ are illustrated as
\begin{align}
H(\varphi_{x}|B) + H(\phi_{x}|B),
\end{align}
which is UPQM. In other words, our approach can also paint a general picture of joint uncertainty between measured system $A$ and quantum memory $B$, which is also explained by the pictures in Fig. 1(b) and Fig. 1(e) from the main text.
 
\subsection*{Appendix B: \; QFGURs $\Rightarrow$ UURs}

In QFGURs, the combinations of probabilities contain all physical information that is accessible for measured systems and quantum memory, which leads to a generalization of UURs~\cite{Friedland2013, Puchala2013, Rudnicki2014}. We prove as follows. Suppose two parties share a physical system $\rho_{AB}=\rho\otimes\rho$, and each of which measures $A_{x}$ and $B_{x}$ (assuming $p(x)$ equals to $1$ for some $x$), respectively. Then we derive the inequalities from Eq.~(\ref{QUR}):
\begin{align}\label{UUR}
\max\limits_{a^{(x)}, \pi} \left\{p\left(a^{(x)} | x\right) 
q\left(\pi\left(a^{(x)}\right) | x\right)\right\}  \leqslant \max\limits_{a^{(x)}, \pi}\zeta^{\text{QFGUR}}_{a^{(x)}} 
&:= \Omega_{1},\notag\\
\max\limits_{i,j, \pi} \left\{ \sum\limits_{a^{(x)}\in\{i,j\}\subset\mathbb{N}_{d}}
p\left(a^{(x)}| x\right) q\left(\pi\left(a^{(x)}\right) | x\right)\right\}
 \leqslant \max\limits_{i,j, \pi} \left\{ \sum\limits_{a^{(x)}\in\{i,j\}\subset\mathbb{N}_{d}}\zeta^{\text{QFGUR}}_{a^{(x)}} \right\} 
 &:= \Omega_{2},\notag\\
&\cdots\notag\\
\max\limits_{ \pi} \left\{ \sum\limits_{a^{(x)}=1}^{d} p\left(a^{(x)}| x\right) q\left(\pi\left(a^{(x)}\right) | x\right)\right\}
\leqslant \max\limits_{ \pi} \left\{ \sum\limits_{a^{(x)}=1}^{d} \zeta^{\text{QFGUR}}_{a^{(x)}} \right\} &:= \Omega_{d}.
\end{align}
Note that Eqs.~(\ref{UUR}) are for a fixed measurement $x$, but choosing the maximum for any one of $d$ outcomes, any two of $d$ outcomes, and all outcomes, respectively. Denoting the probability distribution for each measurement $x$ as $\bm{\mathcal{P}} := \left( p\left(a^{(x)}| x\right) \right)_{a^{(x)}}$ for $A_{x}$ and $\bm{\mathcal{P'}} :=\left( q\left(\pi\left(a^{(x)}\right) | x\right) \right)_{a^{(x)}}$ for $B_{x}$ and defining the state-independent vector
\begin{equation}
\bm{\omega}:=\left(\Omega_{1}, \Omega_{2} - \Omega_{1}, \ldots, \Omega_{d} - \Omega_{d-1} \right) \in {\mathbb{R}^{d}}
\end{equation}
leads to the product-form UURs~\cite{Friedland2013, Puchala2013}: $\bm{\mathcal{P}}\otimes\bm{\mathcal{P'}} \prec \bm{\omega}$, with "$\prec$" standing for majorization \cite{Schur}. In the case illustrated in Fig. 1(c) and Fig. 1(e) of the main text, the approach of UURs is a special case of QFGURs. 

UURs indicate that all nonnegative Schur-concave functions can be used here to measure the uncertainties, and hence the corresponding entropic uncertainty relation reads $H(\bm{\mathcal{P}})+H(\bm{\mathcal{P'}})\geqslant H(\bm{\omega})$, here $H$ stands for Shannon entropy. On the other hand, one can assume that, if $\rho_{AB}=\rho\otimes\frac{1}{d} \mathds{1}$ (here $d$ denotes the dimension of subsystem $A$ ($B$)), then QFGUR degenerates to FGUR up to a scalar. Without loss of generality, let us choose $A_{1}=A_{x}$ and $A_{2}=B_{x}$; then the direct-sum form $\bm{\mathcal{P}} \oplus \bm{\mathcal{P'}}$ \cite{Rudnicki2014} is obtained from FGUR, and these schemes have been shown in Fig. 1(d) and Fig. 1(e) respectively in the main text.

\subsection*{Appendix C: \; Majorization Inequalities}

For completeness we start from the derivation of a majorization inequality: observe that for nonnegative vectors $\bm{P}$, $\bm{Q}$ and $\bm{W}$, i.e. components of the corresponding vector are nonnegative, satisfy $\bm{P} \prec \bm{W}$, then 
$\bm{P} \cdot \bm{Q} \leqslant \bm{W}^{\downarrow} \cdot \bm{Q}^{\downarrow}$. Here the down-arrow notation denotes that the components of the corresponding vector are
ordered in decreasing order.

In order to simplify the proof, we apply rearrangement inequality and obtain
\begin{align}
\bm{P} \cdot \bm{Q} \leqslant 
\bm{P}^{\downarrow} \cdot \bm{Q}^{\downarrow}.
\end{align}
Hence, we only need to prove $\bm{P}^{\downarrow} \cdot \bm{Q}^{\downarrow} \leqslant \bm{W}^{\downarrow} \cdot \bm{Q}^{\downarrow}$. 

At first, we assume the length of vectors is $2$, i.e. $l\left(\bm{P}\right)=l\left(\bm{Q}\right)=l\left(\bm{W}\right)=2$, and express them in the following form
\begin{align}
\bm{P}^{\downarrow}=\left(p_{1}, p_{2}\right),\ \bm{Q}^{\downarrow}=\left(q_{1}, q_{2}\right),\ 
\bm{W}^{\downarrow}=\left(w_{1}, w_{2}\right).
\end{align}
Thus
\begin{align}
\bm{P}^{\downarrow} \cdot \bm{Q}^{\downarrow}
&=p_{1}q_{1}+p_{2}q_{2}\notag\\
&=p_{1}q_{1}+\left(p_{1}+p_{2}-p_{1}\right)q_{2}\notag\\
&=p_{1}\left(q_{1}-q_{2})+(p_{1}+p_{2}\right)q_{2}\notag\\
&\leqslant w_{1}\left(q_{1}-q_{2}\right)+\left(w_{1}+w_{2}\right)q_{2}\notag\\
&=\bm{W}^{\downarrow} \cdot \bm{Q}^{\downarrow}.
\end{align}
Assume it is also true for $l\left(\bm{P}\right)=l\left(\bm{Q}\right)=l\left(\bm{W}\right)=n-1$. Now consider the cases with $l\left(\bm{P}\right)=l\left(\bm{Q}\right)=l\left(\bm{W}\right)=n$, 
\begin{align}
\bm{P}^{\downarrow}&=\left(p_{1}, p_{2}, \ldots, p_{n}\right),\notag\\
\bm{Q}^{\downarrow}&=\left(q_{1}, q_{2} \ldots, q_{n}\right),\notag\\
\bm{W}^{\downarrow}&=\left(w_{1}, w_{2}, \ldots, w_{n}\right),
\end{align}
and then rewrite the product $\bm{P}^{\downarrow} \cdot \bm{Q}^{\downarrow}$ as follows,
\begin{align}
\bm{P}^{\downarrow} \cdot \bm{Q}^{\downarrow}&=\sum\limits_{i=1}^{n}p_{i}q_{i}=\sum\limits_{i=1}^{n-1}p_{i}q_{i}+p_{n}q_{n}\notag\\
&\leqslant\sum\limits_{i=1}^{n-2}w_{i}q_{i}+\left(w_{n-1}+w_{n}-p_{n}\right)q_{n-1}+p_{n}q_{n}\notag\\
&\leqslant\sum\limits_{i=1}^{n-1}w_{i}q_{i}+w_{n}q_{n}=\bm{W}^{\downarrow} \cdot \bm{Q}^{\downarrow}.
\end{align}
Thus, we conclude that
\begin{align}
\bm{P} \cdot \bm{Q} \leqslant 
\bm{P}^{\downarrow} \cdot \bm{Q}^{\downarrow} \leqslant
\bm{W}^{\downarrow} \cdot \bm{Q}^{\downarrow}.
\end{align}

\subsection*{Appendix D: \; Proofs of Theorems \ref{thm1} and \ref{thm3}}
 
In the main text, we give the notations for measurements, outcomes, corresponding probability distributions, entanglement functional and steering functional. Here for measurement $x$, denoting the probability distributions $(p_{\lambda}(a|x))_{a}$ as $\bm{P}_{x}$ and $(q_{\sigma_{\lambda}}(a|x))_{a}$ as $\bm{Q}_{x}$, then
assuming
\begin{align}\label{defprob}
\bm{P} :=\bigoplus\limits_{x}\bm{P}_{x},\quad \bm{Q} :=\bigoplus\limits_{x}\bm{Q}_{x},
\end{align}
and thus we have
\begin{align}
S_{\mathcal{E}}=\sum\limits_{\lambda}p\left(\lambda\right)\left(\bm{P}\cdot \bm{Q}\right),
\end{align}
which is the Eq.~(6) of the main text. Note that both probability vectors $\bm{P}$ and $\bm{Q}$ are functions of local hidden variable $\lambda$.

We proceed by introducing majorization: a vector $x$ is majorized by another vector $y$ in $\mathbb R^n$ : $x\prec y$ if
$\sum\limits_{i=1}^{k}x_{i}^{\downarrow}\leqslant\sum\limits_{i=1}^{k}y_{i}^{\downarrow} (k=1, 2, \ldots, n-1)$
and $\sum\limits_{i=1}^{n}x_{i}^{\downarrow}=\sum\limits_{i=1}^{n}y_{i}^{\downarrow}$,
where the down arrow denotes that the components are
ordered in decreasing order $x_{1}^{\downarrow}\geqslant \cdots \geqslant x_{d}^{\downarrow}$.
A nonnegative Schur-concave function $\Phi$ on $\mathbb R^n$ preserves the partial order in the sense that
$x\prec y$ implies that $\Phi(x)\geqslant \Phi(y)$. We take the conventional expression of a probability distribution
vector in short form by omitting the string of zeroes at the end, for instance, $\left(0.7, 0.3, 0, \ldots, 0\right)=\left(0.7, 0.3\right)$,
and therefor the actual dimension of the vector should be clear from the context.

For the probability distributions which come from quantum state under projective measurements $\{|\phi_{x}^{a}\rangle\langle\phi_{x}^{a}|\}$, it must follow the direct-sum majorization uncertainty relation~\cite{Rudnicki2014}
\begin{align}
\bm{Q}\prec \bm{W}^{B},
\end{align}
with
\begin{align}
\bm{W}^{A}&=\left(1, \mathcal{S}_{1}^{A}-1, \mathcal{S}_{2}^{A}-\mathcal{S}_{1}^{A}, \ldots, \mathcal{S}_{dN-1}^{A}-\mathcal{S}_{dN-2}^{A}\right),\notag\\
\bm{W}^{B}&=\left(1, \mathcal{S}_{1}^{B}-1, \mathcal{S}_{2}^{B}-\mathcal{S}_{1}^{B}, \ldots, \mathcal{S}_{dN-1}^{B}-\mathcal{S}_{dN-2}^{B}\right).
\end{align}

Meanwhile, the classical probability distributions, which may not come from the measurements for quantum state, do not have the restriction from quantum uncertainty relation, and they are only majorized by $\left(1, 0, \ldots, 0\right)$, i.e.
\begin{align}
\bm{P}_{x}\prec \left(1, 0, \ldots, 0\right) ~~\forall x.
\end{align}
Furthermore, for the direct-sum of all $\bm{P}_{x}$, we have
\begin{align}
\bigoplus \bm{P}_{x}\prec\bigoplus\limits_{\text{N $~$times}} \left(1, 0, \ldots, 0\right).
\end{align}

\noindent In particular, denoting $\bm{R}$ as
\begin{align}
\bm{R}:=(\underbrace{1,1, \ldots, 1}_{\text{N $~$times}}, 0, \ldots, 0),
\end{align}
and applying our majorization inequality, one can derive
\begin{align}\label{inq: majorization}
\bm{P}\cdot \bm{Q}\leqslant \bm{R}\cdot \bm{W}^{B}=\mathcal{S}_{N-1}^{B}.
\end{align}
Based on Eq. (\ref{inq: majorization}), we can construct a simple bound for our QFGURs, which can be computed explicitly and used to detect steerability
\begin{align}\label{EPR}
S_{\mathcal{E}}\leqslant\mathcal{S}_{N-1}^{B}.
\end{align}
The violation of Eq. (\ref{EPR}) indicates that the quantum state is steerable from Alice to Bob. Similarly, the state is entangled if the following inequality 
\begin{align}
S_{\mathcal{S}}\leqslant \bm{W}^{A}\cdot \bm{W}^{B}=\mathcal{S}^{AB},
\end{align}
is violated. Here, $\bm{W}^{i}$ ($i=A$ or $B$) stands for the direct-sum majorization bound for the measurements on $i$'s system. We complete the proof of our main theorems.

\subsection*{Appendix E: \; State-dependent Bounds for QFGURs}

In principle, we wish to find a state-independent bound for uncertainty relations which can reveal the incompatibility between observables. However, evidence suggests that incompatibility provides only partial information for steerability. In terms of the eigenvalues of the reduced system, the bounds of both Eqs.~(8) and (10) appeared in the main text can be improved and the asymmetry of EPR steering can be revealed:
\begin{cor}
The EPR steering functional $S_{\mathcal{E}}$ (from Alice to Bob) satisfies
\begin{equation}
\sup\limits_{\rho_{AB} \in \mathcal{E}} S_{\mathcal{E}} \leqslant \mathcal{S}_{N}^{B}(\lambda_{B}).
\end{equation}
\end{cor}
\begin{cor}
The EPR steering functional $S_{\mathcal{E}}$ (from Bob to Alice) satisfies
\begin{equation}
\sup\limits_{\rho_{AB} \in \mathcal{E}} S_{\mathcal{E}} \leqslant \mathcal{S}_{N}^{A}(\lambda_{A}).
\end{equation}
\end{cor}
Note that, unlike our main theorems which are only valid for projective measurements, above corollaries hold for any positive-operator valued measures (POVMs). The formal definitions of $\mathcal{S}_{N}^{A}(\lambda_{A})$, $\mathcal{S}_{N}^{B}(\lambda_{B})$ and the proofs of above corollaries are given below. In actuality, performing the same measurements on each system leads to $\mathcal{S}_{N}^{A}=\mathcal{S}_{N}^{B}$, meanwhile $\mathcal{S}_{N}^{A}(\lambda_{A}) \neq \mathcal{S}_{N}^{B}(\lambda_{B})$ in general. Similarly, we can also improve the entanglement functional. 

Quantum states obey the Heisenberg uncertainty principle, meaning that there is an inherent  ``minimum uncertainty" for incompatible observables. To reveal intrinsic limitations on all quantum states, one need study the bound which is independent of state. For example, let $A_{x}$, $B_{x}$ be POVMs given by
\begin{align}
A_{x} = \left\{\Pi^{A, x}_{a^{(x)}}\right\}_{a^{(x)}\in\mathbb{N}_{N_{x}}},
\quad B_{x} &=\left\{\Pi^{B, x}_{a^{(x)}}\right\}_{a^{(x)}\in\mathbb{N}_{M_{x}}},
\end{align}
which includes the following probability distributions of $\rho$
\begin{align}
\bm{P}_{x} &:= \left( \Tr\left( \Pi^{A, x}_{1}\rho\right), \ldots, \Tr\left( \Pi^{A, x}_{N_{x}}\rho\right)\right),\notag\\
\bm{Q}_{x} &:= \left( \Tr\left( \Pi^{B, x}_{1}\rho\right), \ldots, \Tr\left( \Pi^{B, x}_{M_{x}}\rho\right)\right).
\end{align}
We next construct their joint uncertainty in direct-sum form
\begin{align}\label{pq}
\bm{P}:=\bigoplus\limits_{x}\bm{P}_{x},
\quad\bm{Q}:=\bigoplus\limits_{x}\bm{Q}_{x}.
\end{align}
It it therefore left to show how can we find a bound that is computable for direct-sum majorization uncertainty relations with POVMs. In fact, the result in \cite{Rudnicki2014} is only valid for projective measurements, and here we generalize them to POVMs. First note that, for any $k$ elements in $\bm{P}$, their sum can be bounded by
\begin{align}\label{sumproof}
\sum\limits_{a^{(1)} \in T_{1}}\Tr\left( \Pi^{A, 1}_{a^{(1)}}\rho\right)+\ldots+\sum\limits_{a^{(N_{x})} \in T_{N_{x}}}\Tr\left( \Pi^{A, N_{x}}_{a^{(N_{x})}}\rho\right)
=\Tr\left( (\sum\limits_{a^{(x)} \in T_{x}} \Pi^{A, x}_{a^{(x)}}) \cdot \rho \right)
\leqslant \max\limits_{|T_{1}|+\ldots+|T_{x}|=k} \left\Vert \sum\limits_{a^{(x)} \in T_{x}}\Pi^{A, x}_{a^{(x)}} \right\Vert_{\infty}.
\end{align}
Here, $T_{x}$ are subsets of distinct indices from $\mathbb{N}_{N_{x}}$ or $\mathbb{N}_{M_{x}}$ (based on the measurements we choose), and $\left\Vert \cdot \right\Vert_{\infty}$ denotes the infinity operator norm--which, for positive operators, coincides with the maximum eigenvalues of its argument. By setting $\mathcal{S}_{k}^{A}$ and $\mathcal{S}_{k}^{B}$ as
\begin{align}\label{POVMs}
\mathcal{S}_{k}^{A}&:=\max\limits_{|T_{1}|+\ldots+|T_{x}|=k} \left\Vert \sum\limits_{a^{(x)} \in T_{x}}\Pi^{A, x}_{a^{(x)}} \right\Vert_{\infty},\notag\\
\mathcal{S}_{k}^{B}&:=\max\limits_{|T_{1}|+\ldots+|T_{x}|=k} \left\Vert \sum\limits_{a^{(x)} \in T_{x}}\Pi^{B, x}_{a^{(x)}} \right\Vert_{\infty}.
\end{align}
We now have all ingredients for constructing the bound for the joint probability distributions $\bm{P}$ and $\bm{Q}$, define
\begin{align}
\bm{W}^{A}&=\left(\mathcal{S}_{1}^{A}, \mathcal{S}_{2}^{A}-\mathcal{S}_{1}^{A}, \ldots, 0\right),\notag\\
\bm{W}^{B}&=\left(\mathcal{S}_{1}^{B}, \mathcal{S}_{2}^{B}-\mathcal{S}_{1}^{B}, \ldots, 0\right),
\end{align}
By the construction of $\bm{W}^{A}$ and Eq. (\ref{sumproof}), the sum of first $k$ largest components of $\bm{P}$ in Eq. (\ref{pq}) cannot be greater than $\mathcal{S}_{k}^{A}$, which implies
\begin{align}
\bm{P} \prec \bm{W}^{A}.
\end{align}
To prove $\bm{Q} \prec \bm{W}^{B}$, we use essentially the same argument, construction and proof as in proving $\bm{P} \prec \bm{W}^{A}$. Moreover, a big difference from Eq. (9) in our main text is that Eq. (\ref{POVMs}) established for general POVMs while Eq. (9) only set up for projective measurements. Here both $\bm{W}^{A}$ and $\bm{W}^{B}$ are state-independent, which means they only depends on the incompatibility between observables, but not the quantum state $\rho$.


On the other hand, it is sometimes useful to consider the eigenvalues of the reduced system $\rho_{A}$ and $\rho_{B}$ from bipartite system $\rho_{AB}$, where we will denote their eigenvalues in vector form
\begin{align}
\text{spectrum}\left(\rho_{A}\right)^{\downarrow}&:=\bm{\lambda}^{A},\notag\\
\text{spectrum}\left(\rho_{B}\right)^{\downarrow}&:=\bm{\lambda}^{B},
\end{align}
the down arrow notation denotes that the components of the corresponding vector are ordered in decreasing order. In the following we present upper bounds for direct-sum majorization uncertainty relations in the presence of eigenvalues from reduced system. When using a bipartite system $\rho_{AB}$ and POVMs $B_{x}$, the $k$ largest components of joint probability distributions in Bob's side satisfy the following inequality
\begin{align}\label{sumproof2}
\sum\limits_{a^{(1)} \in T_{1}}\Tr\left( \Pi^{B, 1}_{a^{(1)}}\rho_{B}\right)+\ldots+\sum\limits_{a^{(M_{x})} \in T_{M_{x}}}\Tr\left( \Pi^{A, M_{x}}_{a^{(M_{x})}}\rho_{B}\right)
=\Tr\left( (\sum\limits_{a^{(x)} \in T_{x}} \Pi^{B, x}_{a^{(x)}}) \cdot \rho_{B} \right)
\leqslant\text{spectrum}\left(\sum\limits_{a^{(x)} \in T_{x}}\Pi^{B, x}_{a^{(x)}}\right)^{\downarrow} 
\cdot \bm{\lambda}^{B} 
\end{align}
Quantitatively, by defining the eigenvalues of $\sum\limits_{a^{(x)} \in T_{x}}\Pi^{A, x}_{a^{(x)}}$ and 
$\sum\limits_{a^{(x)} \in T_{x}}\Pi^{B, x}_{a^{(x)}}$ as
\begin{align}
\text{spectrum}\left(\sum\limits_{a^{(x)} \in T_{x}}\Pi^{A, x}_{a^{(x)}}\right)^{\downarrow} &:= 
\bm{\lambda}\left(\sum\limits_{a^{(x)} \in T_{x}}\Pi^{A, x}_{a^{(x)}}\right),\notag\\
\text{spectrum}\left(\sum\limits_{a^{(x)} \in T_{x}}\Pi^{B, x}_{a^{(x)}}\right)^{\downarrow} &:= 
\bm{\lambda}\left(\sum\limits_{a^{(x)} \in T_{x}}\Pi^{B, x}_{a^{(x)}}\right).
\end{align}
and setting $\mathcal{S}_{k}^{A}\left(\lambda_{A}\right)$, $\mathcal{S}_{k}^{B}\left(\lambda_{B}\right)$ as
\begin{align}
\mathcal{S}_{k}^{A}\left(\lambda_{A}\right) &:= \max\limits_{|T_{1}|+\cdots+|T_{x}|=k} 
\bm{\lambda}^{A} \cdot \bm{\lambda}\left(\sum\limits_{a^{(x)} \in T_{x}}\Pi^{A, x}_{a^{(x)}}\right),\notag\\
\mathcal{S}_{k}^{B}\left(\lambda_{B}\right) &:= \max\limits_{|T_{1}|+\cdots+|T_{x}|=k} 
\bm{\lambda}^{B} \cdot \bm{\lambda}\left(\sum\limits_{a^{(x)} \in T_{x}}\Pi^{B, x}_{a^{(x)}}\right).
\end{align}
with 
\begin{align}
\bm{W}^{A}\left(\lambda_{A}\right)&:=\left(\mathcal{S}_{1}^{A}\left(\lambda_{A}\right), \mathcal{S}_{2}^{A}\left(\lambda_{A}\right)-\mathcal{S}_{1}^{A}\left(\lambda_{A}\right), \ldots, 0\right),\notag\\
\bm{W}^{B}\left(\lambda_{B}\right)&:=\left(\mathcal{S}_{1}^{B}\left(\lambda_{B}\right), \mathcal{S}_{2}^{B}\left(\lambda_{B}\right)-\mathcal{S}_{1}^{B}\left(\lambda_{B}\right), \ldots, 0\right).
\end{align}

\noindent Therefore, we have established
\begin{align}
\bm{Q} \prec \bm{W}^{B}\left(\lambda_{B}\right),
\end{align}
for joint probability distributions $\bm{Q}$. Similarly, we can derive the bound for system $A$ with POVMs $A_{x}$ as $\bm{P} \prec \bm{W}^{A}\left(\lambda_{A}\right)$. A natural follow up question concerns the relations between $\bm{W}^{A}\left(\lambda_{A}\right)$, $\bm{W}^{B}\left(\lambda_{B}\right)$, $\bm{W}^{A}$ and $\bm{W}^{B}$. Clearly, by taking $A_{x}$ and $B_{x}$ as projective measurements, we obtain
\begin{align}
\bm{P} \prec \bm{W}^{A}\left(\lambda_{A}\right) \prec \bm{W}^{A},
\quad \bm{Q} \prec \bm{W}^{B}\left(\lambda_{B}\right) \prec \bm{W}^{B},
\end{align} 
which covers the main results appeared in \cite{Puchala2017}. More precisely, we have shown that
our results of $\bm{W}^{A}\left(\lambda_{A}\right)$ and $\bm{W}^{B}\left(\lambda_{B}\right)$ under any POVMs are generalizations of the bound that appeared in recent work \cite{Puchala2017},  which only holds for rank-one projective measurements.

%
%
%
%

We now have all the necessary ingredients for proving Corollary 5. Denoting
\begin{align}
\bm{P}_{x}^{\lambda} &:= \left(p_{\lambda} (a|x)\right)_{a},\notag\\
\bm{Q}_{x}^{\lambda} &:= \left(\Tr(B_{x}^{a} \sigma_{\lambda})\right)_{a},\notag\\
\bm{Q}_{x}^{B} &:= \left(\Tr(B_{x}^{a} \rho_{B})\right)_{a},
\end{align}
we observe that
\begin{align} 
\sum\limits_{\lambda} p_{\lambda} \bigoplus\limits_{x=1}^{N}\bm{Q}_{x}^{\lambda}=\bigoplus\limits_{x=1}^{N}\bm{Q}_{x}^{B}.
\end{align}
Here the `sum' stands for element-wise sum, for example $\left(1, 0\right) + \left(0, 1\right) = \left(1, 1\right)$. Then, according to direct-sum majorization, we can derive
\begin{align}
\bigoplus\limits_{x=1}^{N}\bm{Q}_{x}^{B} \prec \bm{W}^{B} \left(\lambda_{B}\right).
\end{align}

On the other hand, recall steering functional $S_{\mathcal{E}}$
\begin{align}
S_{\mathcal{E}} &=\sum\limits_{\lambda} p\left(\lambda\right) \left( \left(\bigoplus\limits_{x=1}^{N}\bm{P}_{x}^{\lambda}\right) \cdot \left(\bigoplus\limits_{x=1}^{N}\bm{Q}_{x}^{\lambda}\right) \right)\notag\\
&\leqslant \sum\limits_{\lambda} p\left(\lambda\right) \left( \bm{R} \cdot \left(\bigoplus\limits_{x=1}^{N}\bm{Q}_{x}^{\lambda}\right) \right)\notag\\
&\leqslant \bm{R} \cdot \bm{W}^{B} \left(\lambda_{B}\right)\notag\\
&=\mathcal{S}_{N}^{B}\left(\lambda_{B}\right),
\end{align}
where the first equality follows from the majorization inequality which has been proved in next section. The proof of Corollary 5 is complete, and similar method can be applied to the proof of Corollary 6.

Now the connection with the asymmetry of EPR steering: By performing $\left\{A_{x}\right\}$ on both systems, the steering functional from Alice to Bob is bounded by $\mathcal{S}^{A}_{N}\left(\lambda_{A}\right)$, while the steering functional from Bob to Alice is bounded by $\mathcal{S}^{A}_{N}\left(\lambda_{B}\right)$. And they do not equal each other in general due to the effect of eigenvalues from different systems. Since $\lambda_{A} \neq \lambda_{B}$, quantum mechanics allows for the possibility of one-way steering.

\subsection*{Appendix F: \; Linear EPR steering Inequalities}

To show the universality of our method, we construct linear EPR steering inequality based our framework, which is of the form \cite{OneWayPryde,Saunders2010}
\begin{align}\label{linear}
\mathcal{S}_{N} = \frac{1}{N} \sum\limits_{x=1}^{N} a^{(x)} \langle B_{x} \rangle \leqslant \mathcal{B}_{N}.
\end{align}
Here the quantity $\mathcal{S}_{N}$ stands for the steering parameter for $N$ measurement settings and $\mathcal{B}_{N}$ is the shape bound for $\mathcal{S}_{N}$. 

On the other hand, the linear combinations of our QFGURs with coefficients $a^{(x)}$ and $b^{(x)}$, involving statistics collected from an experiment with $N$ measurement settings for each side are
\begin{align}
\left\{ a^{(x)} b^{(x)}\sum\limits_{\lambda} p(\lambda) p_{\lambda}(a^{(x)} | x) q_{\sigma_{\lambda}} (b^{(x)} | x) | \forall \bm{a}, \bm{b} \right\},
\end{align}
clearly it may not be a convex combination of our QFGURs. 
If we make no assumption that Alice's announcement $a^{(x)}$ is derived from a real quantum measurement, the sum of our combinations forms linear EPR steering inequalities is 
\begin{align}
\frac{1}{N} \sum\limits_{x=1}^{N} \sum\limits_{b=1}^{d}\sum\limits_{\lambda} a^{(x)} b^{(x)} p\left(\lambda\right) p_{\lambda}\left(a^{(x)} | x\right) q_{\sigma_{\lambda}} \left(b^{(x)} | x\right)
= \frac{1}{N} \sum\limits_{x=1}^{N} a^{(x)} \langle B_{x} \rangle_{\sigma_{\lambda}},
\end{align}
where $B_{x}=\sum\limits_{b} b B_{x}^{b}$. With this, we rewrite the left hand of Eq. (\ref{linear}).

Demonstrating the product of accommodation coefficients, i.e. $a^{(x)} b^{(x)}$, in decreasing order and denote it as $\bm{\mathcal{C}}$
\begin{align}
\bm{\mathcal{C}}=\left(a^{(x)} b^{(x)}\right)^{\downarrow},
\end{align}
and the length of vector $\bm{\mathcal{C}}$ is $Nd$ since for each measurement $x$, Alice only declares one outcome $a^{(x)}$ while $b^{(x)}$ has $d$ different possibilities in each round of the measurement. Now it is easy to construct a bound from our QFGUR
\begin{align}\label{linear bound}
\mathcal{S}_{N} \leqslant \frac{\bm{\mathcal{C}} \cdot \bm{W}^{B}}{N},
\end{align}
simply from the fact $0 \leqslant p_{\lambda}\left(a^{(x)} | x\right) \leqslant 1$. Any violation of Eq. (\ref{linear bound}) implies the steerability from Alice to Bob. 

\subsection*{Appendix G: \; Examples}

\textit{Examples.}--Werner states~\cite{Wernerstate} are the best-known class of mixed entangled state. In the following, we consider various families of Werner states and show some numerical results to compare our criterion with previous works.

(i)~$2\times2$ Werner states.--
First, let us consider $\rho_{w}=p|\phi^{-}\rangle\langle\phi^{-}|+(1-p)\mathds{1}/4$, where $p\in[0,1]$ and $\ket{\phi^{-}}$ is the singlet state. To test its steering and entanglement with Eqs.~(8) and (10) obtained in our main text, we choose two spin measurements $\sigma_x$ and $\sigma_z$, which give us $S_{\mathcal{Q}}=1+p$, $\mathcal{S}_{0}^{B(A)}=1$, $\mathcal{S}_{1}^{B(A)}=1+1/\sqrt{2}$ and $\mathcal{S}_{2}^{B(A)}=\mathcal{S}_{3}^{B(A)}=2$, for $N=2$ and $d=2$. Therefore, the Werner state is steerable when $p>1/\sqrt{2}$ based on criterion (8), which coincides with the previous result \cite{Howard07PRL}. For three spin measurements $\sigma_x$, $\sigma_y$ and $\sigma_z$ we find that based on our framework the Werner states is steerable whenever $p>1/\sqrt{3}$, which coincides with the previous optimal result again~\cite{Saunders2010}. On the other hand, the state is entangled when $p>2-\sqrt{2}$ based on criterion (10). It has been proven that two-qubit Werner states are entangled if and only if (iff) $p>1/3$ \cite{Wernerstate}, so our condition for entanglement is sufficient but not necessary.

(ii)~$3\times3$ Werner states.--
We also consider a high dimensional case for qutrits with $\rho_{w}=p|\psi^{+}\rangle\langle\psi^{+}|+(1-p)\mathds{1}/9$, where $|\psi^{+}\rangle=(|00\rangle+|11\rangle+|22\rangle)/\sqrt{3}$. For the following three Gell-mann measurements
\begin{equation}
\lambda_{1}=
\left(
  \begin{array}{ccc}
    0 & 1 & 0 \\
    1 & 0 & 0 \\
    0 & 0 & 0 \\
  \end{array}
\right),~
\lambda_{4}=
\left(
  \begin{array}{ccc}
    0 & 0 & 1 \\
    0 & 0 & 0 \\
    1 & 0 & 0 \\
  \end{array}
\right),~
\lambda_{8}=\frac{1}{\sqrt{3}}
\left(
  \begin{array}{ccc}
    1 & 0 & 0 \\
    0 & 1 & 0 \\
    0 & 0 & -2 \\
  \end{array}
\right),
\end{equation}
we have $S_{\mathcal{Q}}=1+2p$, $\mathcal{S}_{0}^{B(A)}=1$, $\mathcal{S}_{1}^{B(A)}=2$, $\mathcal{S}_{2}^{B(A)}=(3+\sqrt{5})/2$ and $\mathcal{S}_{3}^{B(A)}=\cdots=\mathcal{S}_{8}^{B(A)}=3$, for $N=3$ and $d=3$.
Therefore, our criteria are efficient to classify a two-qutrit Werner state being steerable when $p>0.809$ and entangled when $p>0.763$. However, if we turn to the method introduced in Ref.~\cite{Rutkowski2017}, we find the right hand side of Eq.~(13) appeared in the main text is $3$ and a trivial condition of $p>1$ to verify steering, which means that method is ineffective in this case. Here in order to derive the quantity $\mathcal{S}_{k}$ while Gell-Mann matrix $\lambda_{8}$ is degenerate, one should choose the eigenvectors that maximize $\mathcal{S}_{k}$. We remark that our method in detecting entanglement and EPR steering depends on the choice of measurements and the combinations of measurement outcomes, which means in order to derive the most efficient bounds, we need to check all possible measurements. However, when measurements are given, our method is stronger than previous result based on uncertainty relations to some extent, as shown in our example (i) and (ii).

\subsection*{Appendix H: \; Generalized QFGURs}

There is a major limitation for our QFGURs defined in the main text, that is, it is only valid for bipartite quantum state $\rho_{AB}$ whose reduced systems have the same dimension, i.e. $\text{dim}(\mathcal{H}_{A})=\text{dim}(\mathcal{H}_{B})$. And here we consider in detail the case in which 
         $\text{dim}(\mathcal{H}_{A}) \neq \text{dim}(\mathcal{H}_{B})$. Without loss of generality, we can assume that 
\begin{align}
\text{dim}(\mathcal{H}_{A})&:=d_{1},\notag\\
\text{dim}(\mathcal{H}_{B})&:=d_{2},
\end{align}
where $d_{2}>d_{1}$. By applying projective measurements $A_{x}$ and $B_{x}$, we can collect their probability distributions as 
\begin{align}
\bm{P}_{x} &:= \left( p(1^{(x)}|x), p(2^{(x)}|x), \ldots, p(d_{1}^{(x)}|x)\right),\notag\\
\bm{Q}_{x} &:= \left( q(1^{(x)}|x), q(2^{(x)}|x), \ldots, q(d_{2}^{(x)}|x)\right).
\end{align}
Note that these two probability distributions have different length, which means that the inner product between them has not been defined, hence no QFGURs can be constructed between them directly, and a modification is needed. In such case we set vector $\overline{\bm{P}}_{x}$ as
\begin{align}
\overline{\bm{P}}_{x} =\left( \overline{p}(1^{(x)}|x), \overline{p}(2^{(x)}|x), \ldots, \overline{p}(d_{2}^{(x)}|x)\right)
:= ( p(1^{(x)}|x), p(2^{(x)}|x), \ldots, p(d_{1}^{(x)}|x), \underbrace{0, \ldots, 0}_{\text{$d_{2}-d_{1}$ $~$times}} ),
\end{align}
with $\overline{p}(a^{(x)}|x)=p(1^{(x)}|x)$ for $a \leqslant d_{1}$, otherwise $\overline{p}(a^{(x)}|x)=0$. The definition of $\overline{\bm{P}}_{x}$ implies that now the QFGURs for $\overline{\bm{P}}_{x}$ and $\bm{Q}_{x}$ are well-defined
\begin{equation}
\overline{U}_{\text{QFGUR}}:=\left\{\sum\limits_{x=1}^{N}p\left(x\right) \overline{p}\left(a^{(x)}| x\right) q\left(\pi\left(a^{(x)}\right) | x\right)\leqslant \zeta^{\text{QFGUR}}_{a}|\forall \textbf{a}, \pi \right\},
\end{equation}
Finally, by defining $\overline{U}_{\text{QFGUR}}$ as the QFGURs for probability distributions $\bm{P}_{x}$ and $\bm{Q}_{x}$, we immediately obtain the generalized QFGURs for asymmetric systems. Moreover, we can also extend this handing mechanism to POVMs easily. 

In order to construct the QFGURs for different measurement settings, for example Alice chooses one of her measurement settings $x\in \mathbb{N}_{N_{1}}$ while Bob chooses measurements $x\in \mathbb{N}_{N_{2}}$ where $\mathbb{N}_{N}:=\left\{1, \ldots,  N\right\}$ and $N_{1} \neq N_{2}$, we can follow a similar construction as shown in the case $d_{1} \neq d_{2}$. This means that we can always assume that $d_{1} = d_{2}$ and $N_{1} = N_{2}$, otherwise we can follow above steps. On the other hand, when Alice chooses one of her measurement settings $x\in \mathbb{N}_{N}$ while at the same time Bob chooses measurements $y\in \mathbb{N}_{N}$, then we need to find a permutation $\pi'$ between $x$ and $y$ such that $y=\pi'(x)$, which establish the following QFGURs
\begin{align}
\overline{\overline{U}}_{\text{QFGUR}}
:=\left\{\sum\limits_{x=1}^{N}p\left(x\right) p\left(a^{(x)}| x\right) q\left(\pi\left(a^{(x)}\right) | \pi'(x)\right)\leqslant \zeta^{\text{QFGUR}}_{a}|\forall \textbf{a}, \pi, \pi' \right\}.
\end{align}
From above considerations, we find that the QFGURs constructed in our main text can be generalized to the cases with asymmetric bipartite systems, different measurements settings or even different choices of measurements.

\end{document}